\def\year{2020}\relax
\documentclass[letterpaper]{article} 
\usepackage{aaai20}  
\usepackage{times}  
\usepackage{helvet} 
\usepackage{courier}  
\usepackage[hyphens]{url}  
\usepackage{graphicx} 
\usepackage{graphicx}  
\title{Multi-Agent Reinforcement Learning and Human Social Factors in Climate Change Mitigation}
\author{
\Large \textbf{Kyle Tilbury, Jesse Hoey}\\ 
David R. Cheriton School of Computer Science\\
University of Waterloo\\
Waterloo, Ontario, Canada\\
ktilbury@uwaterloo.ca, jhoey@cs.uwaterloo.ca 
}

\begin{document}
\maketitle
\begin{abstract}
    Many complex real-world problems, such as climate change mitigation, are intertwined with human social factors. Climate change mitigation, a social dilemma made difficult by the inherent complexities of human behavior, has an impact at a global scale. We propose applying multi-agent reinforcement learning (MARL) in this setting to develop intelligent agents that can influence the social factors at play in climate change mitigation. There are ethical, practical, and technical challenges that must be addressed when deploying MARL in this way. In this paper, we present these challenges and outline an approach to address them. Understanding how intelligent agents can be used to impact human social factors is important to prevent their abuse and can be beneficial in furthering our knowledge of these complex problems as a whole. The challenges we present are not limited to our specific application but are applicable to broader MARL. Thus, developing MARL for social factors in climate change mitigation helps address general problems hindering MARL's applicability to other real-world problems while also motivating discussion on the social implications of MARL deployment.
\end{abstract}

\noindent Real-world issues affecting humanity can be heavily influenced by social factors and, consequently, can be influenced by software agents. This has been demonstrated by past work on aiding homeless shelters using software agents. Such agents were used to help effectively spread awareness about HIV preventative practices among homeless youth by strategically using the social network of the target population to maximize information exchange~\cite{yadav2016using}. Another relevant issue impacted by social factors and software agents is voting. The structure of a social network can influence voting decisions and bots inserted into the network can be used to achieve certain election outcomes~\cite{stewart2019information}. Climate change, which is an issue that has an impact at a global scale, is also affected by social factors. This is shown by socio-climate models which are models that couple the geophysical aspects of climate change with human social factors. These models capture how human behavior can impact climate change and how, in turn, climate change can affect human behavior. Two impactful factors are social learning, which is the process whereby individuals learn new behaviors, values and opinions from others, and social norms, which are socially accepted and widely practiced modes of conduct~\cite{bury2019charting}. It is not yet clear how intelligent software agents could influence human behaviors and the social factors at play in climate change mitigation.

We propose developing multi-agent reinforcement learning (MARL) based agents that are capable of influencing human social factors within the climate change mitigation setting. Multi-agent reinforcement learning is commonly demonstrated with social dilemmas~\cite{leibo2017multi,peysakhovich2018prosocial,tampuu2017multiagent,jaques2019social}. These are typically toy \emph{tragedy of the commons} type problems where individual agents can obtain high reward in the short term by exploiting reward mechanisms, but in the long term every agent involved achieves a higher reward when they cooperate. Climate change mitigation is ultimately a social dilemma as well~\cite{milinski2008collective}. Specifically, climate change can be characterized as a social dilemma wherein: people have to make repeated decisions over relatively long timescales before the outcome is evident, the value of mitigating climate change can be non-obvious, the investments people make towards mitigation are not necessarily recouped, and there are potentially disastrous consequences if a threshold of mitigation is not met~\cite{milinski2008collective}.

Applying MARL to real-world climate change mitigation is more difficult than applying it to simple social dilemmas, where there are typically only a few different types of agents, due to the abundant types of humans with varying attitudes, motivations, behaviors, and capacities to affect climate change mitigation. For example, a CEO of a multi-national company could have a large impact on climate change but is relatively unaffected by it, whereas a simple farmer could have very little impact on climate change but is highly susceptible to the effects. MARL agents could take these disparities into account to best direct mitigation efforts. Furthermore, MARL agents could allow for the complexities of human behaviors. These complexities can lead to unintended outcomes as demonstrated when simply communicating mean energy usage to consumers. This can lead to over-users reducing their usage, but also can cause an unintended ``boomerang'' effect where under-users increasing their usage~\cite{schultz2007constructive}. However, the addition of an emotional signal (a positive emoji to the under-users) can reduce this boomerang effect. A team of MARL agents operating on a social network could request usage information from each other, discover the mitigatory effect of communicating this information to humans and the resulting boomerang effect, and then further discover the resolution provided by simple emotional signaling. Research into such agents needs to address ethical concerns, practical implementation issues, and technical problems relating to interacting with noisy and unpredictable humans.

\section{Problems}
We discuss potential problems that arise from applying MARL to the proposed area of social factors affecting climate change mitigation. The potential issues we identify fall under three main categories: ethical, practical, and technical. These issues are not limited to our specific application, but also have implications to broader MARL.

\subsection{Ethical}
First, we consider the ethical and moral concerns inherent to intelligent agents capable of influencing humans socially. It seems inevitable that these types of agents will come to exist, which gives rise to important questions. How should agents that can impact human social factors be used? If these agents are learning from a human's personal information to better influence them, then there are inherent privacy issues. If these agents are adept at changing human behavior, then there are clear issues regarding human autonomy and ability to make choices for themselves. Furthermore, what are the real-world implications and ramifications of socially intelligent agents? Of particular importance are potential biases that these agents could create or propagate. With regard to climate change mitigation, one could imagine an oil corporation using intelligent agents to act against climate change mitigation. Alternatively, one could also imagine a solar panel company acting for climate change mitigation, but tending toward strategies that increase the sale of solar panels even though other more effective overall strategies may exist. More broadly, intelligent agents capable of influencing humans socially could be used to spread any number of harmful ideas or propaganda. Will more good come from this type of agent than harm? Given the cataclysmic effects that result from climate change and the potential for these agents help mitigate these effects, it is worthwhile to consider the existence of such agents despite the potential for misuse. Finally, who should decide on what constitutes ethical use of MARL agents in this setting and who will ultimately be responsible for them? Addressing these and other questions and understanding how intelligent agents can operate on social factors in the real world will enable us to ascertain the potential risks of such agents and prevent their abuse.

\subsection{Practical}
Second, there are problems associated with the practical nature of implementing MARL based agents for influencing human social factors in climate change mitigation. One possible implementation would be to have the agents interact with humans as intelligent personal assistants (e.g. in their smartphone). An intelligent personal assistant could track a user's beliefs and behaviors and even model their mental state~\cite{gmytrasiewicz2018interactive}. This could be valuable to an agent in the climate change mitigation setting where having access to this kind of information and a long term objective could enable the agent to learn an individualized strategy for that specific human. Another possible implementation is agents that are inserted into and act in social networks (e.g. in Twitter). Determining people's climate change mitigation behaviors and beliefs within the social network would be a difficult task. The actions of agents in this setting would have to be carefully directed and carried out. Recall the example where a message about the social norms of average power consumption sent to consumers may either lower usage, in the case of above average consumers, or have a boomerang effect and increase usage among below average consumers~\cite{schultz2007constructive}. Additionally, different people within the same group can react differently to the same message. So, carefully communicating with humans is necessary to have the desired effects. Understanding how to design systems that allow for effective communication among agents and humans is an important practical aspect that must be addressed~\cite{crandall2018cooperating}.

\subsection{Technical}
Third, we consider the technical problems. These problems not only apply to our proposed area but are also directly applicable to broader MARL. Agents socially influencing humans in an environment may be limited to act based only on locally observable information. Additionally, the actions they take may start at, stop at, and last for varying timelines. These are challenges that affect our problem, but also previous MARL applications~\cite{amato2019modeling}. However, there are further questions that will need to be addressed due to the human component of our problem. There can be long time scales time before social changes manifest which is made more difficult to account for due to the asynchronous nature of agent's actions. What strategies would allow for agents to learn under this limitation? Environment states and rewards are affected by the joint actions of all the agents and the humans. How can agents learn robust strategies while learning in an environment that is dynamic and potentially non-stationary due to the human elements? Further, how can MARL agents learn human preferences or cost functions adequately to be able to accurately predict human actions?

It is necessary to address both how agents treat one another and how agents treat humans within the environment. Agents may be oblivious of each other, treating other agents simply as part of the environment~\cite{lanctot2017unified,leibo2017multi}, or they may act as a team with the same ultimate goal~\cite{amato2019modeling}. In the case of humans, there are also different ways that agents could treat them. Humans could be regarded as part of the environment that the agents are operating in, requiring the agents to learn models of them. The learning process could be bootstrapped and the agents given a model of humans. The complexity of this model of humans and their behaviors will have to be specified. Models that take into account social and emotional factors in human decision making may be important in this regard~\cite{hoey2016affect}.

There is a need to identify possible reward mechanisms for achieving desired behavior of the agents. Social influence has been used as a reward mechanism in previous MARL settings to achieve coordination and communication among agents~\cite{jaques2019social}. How could agents use social influence as reward in settings that involve humans alongside other agents? Different reward mechanisms that target different social factors could result in different emergent behaviors. As social learning is highly relevant in mitigation and adaptation to climate change, it is desirable to identify reward mechanisms where the emergent behavior speeds up or slows down social learning in humans. Similarly, rewards that result in emergent behaviors that influence human behavior regarding social norms should also be investigated.

Another aspect of the problem is the interplay between agents with incompatible goals trying to influence the same humans. With voting, different groups of artificial agents acting towards goals that are diametrically opposed to one another can have negative effects on the entire voting process~\cite{stewart2019information}. Consider climate change mitigation, there could be a pro-climate change mitigation group of agents and an anti-climate change mitigation group of agents all attempting to socially influence humans. Do opposing groups of intelligent agents acting on the same network of individuals have similar pervasive negative effects regarding climate change mitigation?

\section{Approach}
We now discuss a high-level approach to employ MARL for social factors in climate change mitigation. We propose developing a socio-climate model that will capture the most important aspects of social factors and climate dynamics. Utilizing a simple model will allow us to easily assess processes and feedback within the system~\cite{bury2019charting}. This climate model will be the basis environment in which to experiment with MARL agents, beginning with a simulation. 

Individuals within the simulation will be simulated humans and agents. We propose to bootstrap the learning process of the agents by providing them with a model of humans regarding climate change mitigation. There could be different models of behavior for different groups of humans in the real world. Some groups could be fanatical, i.e. \emph{mitigators} who would never become \emph{non-mitigators} and non-mitigators who would never become mitigators. Other groups comprised of mitigators and non-mitigators that can be reasonably swayed to change their alignment, which we call \emph{swayable}, could have more interesting qualities. It is this swayable group that we would seek to model and to understand the dynamics of (i.e. how do they update themselves based on social and geophysical factors). One potential approach could be to identify and remove the fanatics by detecting their extreme stances on climate change mitigation in tweets~\cite{mohammad2017stance}. People within groups will react differently to different factors, such as the groups average power consumption, affecting climate change~\cite{schultz2007constructive}. The MARL agents could learn how to address these factors intelligently and appropriately. We would seek to learn the model and reward functions that humans are optimizing with regard to climate change mitigation behaviors. Inverse reinforcement learning could be a possible way of achieving this~\cite{abbeel2004apprenticeship}. Based on the learned swayable human model, the simulated humans will behave as mitigators or non-mitigators and can switch to the more attractive mode of behavior based on the factors that were learned from real world dynamics. Additionally, we will model the mechanisms that shape groups of humans to decide on a collective action such as rewards and sanctions~\cite{gois2019reward}.

Multiple types of intelligent agents will also be investigated. These agents will be \emph{mitigation influencers}, which attempt to influence humans to become mitigators, or, similarly, \emph{non-mitigation influencers}. This will build an intuition of the effect of intelligent agents on the network dynamics and simulated humans. Examining different reward mechanisms, such as social influence~\cite{jaques2019social} or esteem~\cite{moutoussis2014formal}, in the cooperative and competitive environment between the different types of agents can allow for different emergent behaviors. Additionally, competition between agents can allow for more robust strategies overall~\cite{tampuu2017multiagent}.  

The intuitions gained from the simulation will be used as a basis to develop an experiment that involves agents interacting with human subjects. A small-scale experiment, with tens of humans, can determine dynamics that scale up and have implications at a global scale~\cite{milinski2008collective}. This kind of experiment will also enable addressing the, both practical and technical, issues of the communication between agents and humans.

We envision the use of intelligent agents on human social factors in climate change mitigation as a means to come to a greater understanding of the underlying systems, rather than as a means to exert control over them. We endeavor to further the understanding of: climate change mitigation strategies, the social factors at play in climate change mitigation, and the ability of intelligent agents to act appropriately in complex social dilemmas. Confronting the ethical issues associated with our approach is not an easy task. The outlined ethical problems are not limited to the future or to our proposed application; they apply to current AI in general. The Montreal Declaration for a Responsible Development of Artificial Intelligence, and its entire set of responsible AI declaration points, brings awareness to the principles that should held to develop ethical and responsible AI technologies~\cite{montreal2017}. We will account for these principles while developing our approach for MARL and human social factors. For the ethical issues relating to privacy and maintaining human autonomy, there could be a direct trade off between efficacy of the approach and the involved human's rights which will be considered. Furthermore, it is important for the actions taken by MARL agents to be transparent and interpretable so that they can be scrutinized. The ethical use of and the responsibility for technologies like MARL begin with those of us who develop them. Ultimately it will be a collaborative effort between society at large, technologists, and legal entities to form a consensus on and enforce the ethical use of such agents.

\section{Conclusion}
Attempting to address human social factors in climate change mitigation through the use of MARL is complex and has implications for society. We identified ethical, practical, and technical problems that should be addressed and outlined a possible approach to begin to mitigate them. The technical issues faced by our application also affect the applicability of MARL in other settings. Therefore, by applying MARL to this application we will address the broader problems facing MARL and motivate solving these problems to a larger set of researchers. Exploring the effects of MARL on the human social factors in climate change mitigation can also increase our understanding of these social factors in general. Evaluating MARL in this setting can highlight the potential quandaries of agents acting within and having effects on human social networks at large. In the long term, understanding how intelligent agents can socially coordinate, communicate, and cooperate with humans will be beneficial in tackling real-world problems at a global scale while minimizing potential downsides.

\bibliography{bibfile}
\bibliographystyle{aaai}
\end{document}